\documentclass{appolb}
\usepackage{epsfig}
\usepackage{eurosym} 
\usepackage{color}
\usepackage{amsmath} 
\usepackage{amssymb} 
\usepackage{dsfont} 
\usepackage{enumerate} 

\newcommand{\R}{\mathbb{R}}

\newcommand{\dd}{\mathrm{d}}
\newcommand{\Dp}{\partial}

\newcommand{\li}{\left}
\newcommand{\ri}{\right}

\newcommand{\cen}[1]{\begin{center} #1 \end{center}}

\newcommand{\dis}{{\rm dis}}

\begin{document}
\title{Extended soft-wall model for the QCD phase diagram
\thanks{Presented at Critical Point and Onset of Deconfinement 2016, Wroc\l aw.}%
}
\author{R. Z\"ollner,
F. Wunderlich,
B. K\"ampfer
\address{Helmholtz-Zentrum Dresden-Rossendorf, PF 510119, D-01314 Dresden, Germany}
\vspace{-0.2cm}
\address{Institut f\"ur Theoretische Physik, TU Dresden, D-01062 Dresden, Germany}}

\maketitle
\begin{abstract}
The soft-wall model, emerging as bottom-up holographic scenario anchored in the AdS/CFT correspondence, displays the disappearance of normalisable modes referring to vector mesons at a temperature $T_{\dis}$ depending on the chemical potential $\mu$, $T_{\dis}(\mu)$. We explore options for making $T_{\dis}(\mu)$ consistent with the freeze-out curve $T_{\rm f.o.}(\mu)$ from relativistic heavy-ion collisions and the cross-over curve $T_{\rm c}(\mu)$ from QCD at small values of $\mu$.
\end{abstract}
  
\section{Introduction}
In lacking still a convincing top-down approach from string theory to a proper gravity dual of QCD one must resort to bottom-up models which are designed to mimic certain wanted features of QCD. Among such approaches is the soft-wall model \cite{KKSS} as a particular realisation of the AdS/CFT correspondence w.r.t. the hadron spectrum, especially vector mesons. While being a phenomenological set-up, the original soft-wall model \cite{KKSS} can be modified to accommodate the Regge type spectrum of radial excitations of selected hadron species at vanishing temperature $T$ and chemical potential $\mu$. Extending the model further to $T>0$ one finds that, at temperatures $T \geq T_{\dis}$, hadrons as normalisable states disappear \cite{Colangelo}. It is tempting to consider such a scenario as an emulation of deconfinement. As shown in \cite{ich}, one can tune the model further to achieve $T_{\dis}=T_{c}$, where $T_c \approx 150$ MeV is the cross-over temperature known from lattice QCD evaluations for 2+1 flavours with physical quark masses. There are options to let disappear all hadron states at $T_{\dis}$ (instantaneous disappearance) or only the ground state, and excited states already disappeared in a narrow corridor below $T_c$ (sequential disappearance). For steering these details the Hawking-Page transition is a central issue. \\
Reference \cite{ich} focused on purely thermal effects. Here, we investigate the options for $T_{\dis}(\mu)$. We provide a special modification of the soft-wall model such to make $T_{\dis}(\mu)$ consistent with $T_{\rm f.o.}(\mu)$ and $T_c(\mu)$, where ``f.o.'' labels the chemical freeze-out and ``c'' is for the cross over. The dependence of $T_{\rm f.o.}$ on $\mu$ is determined nowadays from hadron multiplicities observed in relativistic heavy-ion collisions at varying beam energy; system size and centrality dependencies help to consolidate the freeze-out curve $T_{\rm f.o.}(\mu)$. The map of hadron multiplicities on the freeze-out data is provided traditionally by thermo-statistical models of the hadron resonance gas \cite{Stachel2, CKWX, Becattini}, may be supplemented by effects of inelastic, post-hadronization reactions \cite{1605.09694}. The results are in agreement with data analyses using mean and variance of net-baryon number and net-electric charge distributions based on lattice QCD input \cite{Bazavov}. On the other hand, lattice QCD provides ab initio calculations of $T_c(\mu)$, albeit restricted to a region $\mu/T <3$ due to the sign problem.\\
All these attempts have the goal to pin down the QCD phase diagram and to seek for a critical point that marks the onset of a curve of first-order phase transitions when going to larger values of $\mu$, realised experimentally by lowing the beam energy. Experimentally, dedicated efforts are devoted to the search for a critical point, most notably the beam energy scan at RHIC \cite{RHIC1, RHIC3, RHIC7} and the program at NA61/SHINE \cite{NA611, NA612, NA613}. Besides many models envisaging statements on the phase diagram of strong interaction matter \cite{Falk, Schaffner1, Schaffner2, Schaffner3, Schaffner4, Schaffner5}, also holographic approaches are to be mentioned. These aim essentially at mimicking the thermodynamics \cite{Gubser, Noronha}, rather than individual hadron properties, but can address issues of deconfinement as well \cite{Kir1, Kir2}. 
Here, we report on the modified soft-wall model with regard of non-zero temperature and non-zero chemical potential. 

\section{Modified soft-wall model} \label{sec2}
The model pursued here is based on the action 
 \begin{equation}
 S_V = -\frac{1}{4k_V} \int  \! \dd z\, \dd^4 x \,   \sqrt{g} e^{-\Phi(z)} F^2  \label{wirkung}
 \end{equation}
with $k_V$ is to be chosen to render $S_V$ dimensionless. The dilaton field $\Phi$ acts as a conformal symmetry breaker. The quantity $g$ denotes the determinant of the metric tensor. Equation (\ref{wirkung}) is utilised to describe the dynamics of an $U(1)$ vector field with the components $V_M$, where $F_{MN} = \partial_M V_N - \partial_N V_M$ (indices $M, N = 0, \ldots, 4$) is the field strength tensor, as dual to the boundary vector current, e.g. $J_{\mu} \sim \bar q \gamma_{\mu} q$. A special five dimensional Riemann space with coordinates $x_{0,1,2,3}$ and holographic coordinate $z$ is described by the infinitesimal distance squared
 \begin{equation}
 d s^2 =  e^{A(z)} \left( f(z) d t^2 - d \vec x ^{\, 2} -\frac{1}{f(z)} d z^2 \right),  \label{ds}
 \end{equation}
where $A(z)$ is a warp function and $f(z)$ is the blackness function, both ones to be specified below. The equation of motion follows from (\ref{wirkung}), with the metric determinant to be read off (\ref{ds}) and $\psi=\varphi \exp\{-(A-\Phi)/2\}$, as
	\begin{equation} 
	\li(\Dp_{\xi}^2 -(U_T-m_n^2) \ri) \psi=0,  \label{schr}
	\end{equation}
where $\xi$ is the tortoise coordinate determined by $\dd \xi = \dd z /f(z)$ and $U_T$ is the Schr\"odinger equivalent potential 
        \begin{equation}
	U_T = \li(\frac12 (\frac12 \Dp_{z}^2 A-\Dp_{z}^2 \Phi) +\frac14 (\frac12 \Dp_{z} A-\Dp_{z} \Phi)^2 \ri) f^2+ \frac14 (\frac12 \Dp_{z} A-\Dp_{z} \Phi) \Dp_{z} f^2. \label{hotpot}
	\end{equation}
To arrive at (\ref{schr}) the ansatz $V_\mu = \epsilon_\mu \varphi(z) \exp\{ i p_{\nu}  x^{\nu} \}$ and the gauges $V_z=0$ and $\Dp_{\mu} V^{\mu} =0$ (Greek indices run in the range $0, \ldots, 3$) are employed. The normalisable solutions of (\ref{schr}) determine squared vector meson masses $m_n^2 = p_{M} p^{M}$, where $n=0$ denotes the ground state (g.s.) and $n\geq1$ counts the radial excitations, labelled by 1$^{\rm st}$, 2$^{\rm nd}$ etc.\\
In the spirit of \cite{KKSS}, the soft-wall model sets a ``soft wall'' by the dilaton profile $\Phi(z) = (cz)^p$ with a scale $c$; we employ the warp factor $A(z) = \ln (L^2/z^2 + \tilde \mu^2 )$ with the AdS radius $L=1/c$. Our ansatz for the blackness function is with $\vartheta(z_H)=\pi z_H T(z_H)-1$ (see Appendix \ref{anhang})
  \begin{equation}
   f(z) = 1-\frac{z^4}{z_H^4} \li(1+ \frac{2\vartheta(z_H)}{\exp\{\frac2e \vartheta(z_H)+4\hat \mu^2\}} \li[ \li(\frac{z}{z_H} \ri)^{2 \exp\{\frac2e \vartheta(z_H)+4\hat \mu^2\}}-1\ri] \ri) \label{f}
  \end{equation}
providing from $\Dp_z f(z) \mid_{z=z_H} = -4\pi T(z_H)$ the Hawking temperature
  \begin{equation}
   T(z_H) = \tilde T(z_H) (1-\hat \mu^2) \label{T}
  \end{equation}
with $\tilde T(z_H)=\tilde T_{\min} (1+[1/x-2+x]/\Theta)$, where $x=z_H/\tilde z_{\min}$ and $\Theta=\pi \tilde T_{\min} \tilde z_{\min}$.
In the special case $\tilde T(z_H)=1/(\pi z_H)$, (\ref{f}) belongs to the metric of a Reissner-Nordstr\"om black hole embedded in an asymptotic Anti-de Sitter space. It is customary to identify $\mu=\sqrt{2}\hat \mu \gamma z_H^{-1}$ as baryo-chemical potential and $T$ as the temperature of the boundary theory. The parameter $\gamma$ arises as ratio of two coupling strengths when deriving the AdS Reissner-Nordstr\"om black brane (cf. \cite{Erdmenger} and Appendix \ref{anhang}). Equation (\ref{f}) keeps the required properties of a black hole: it has a simple zero at horizon $z=z_H$, $f(z=0,z_H) = 1$ and $(\Dp^i_z f)_{z \to 0} =0 $ for $i=1,2,3$. The above parameters $c$, $p$ and $\tilde \mu$ can be tuned at $T=0$ to reproduce a Regge type spectrum $m_n^2=\alpha + \beta n$ in agreement with known vector meson states forming a trajectory of radial excitations parametrised by $\alpha$ and $\beta$ \cite{ich}. Note that, for $\mu=0$, Eq.~(\ref{f}) facilitates numerical results agreeing on the sub percent level with those of \cite{ich}.

\section{Non-zero chemical potential} \label{sec3}
\begin{figure}
 \cen{\includegraphics[scale=.49]{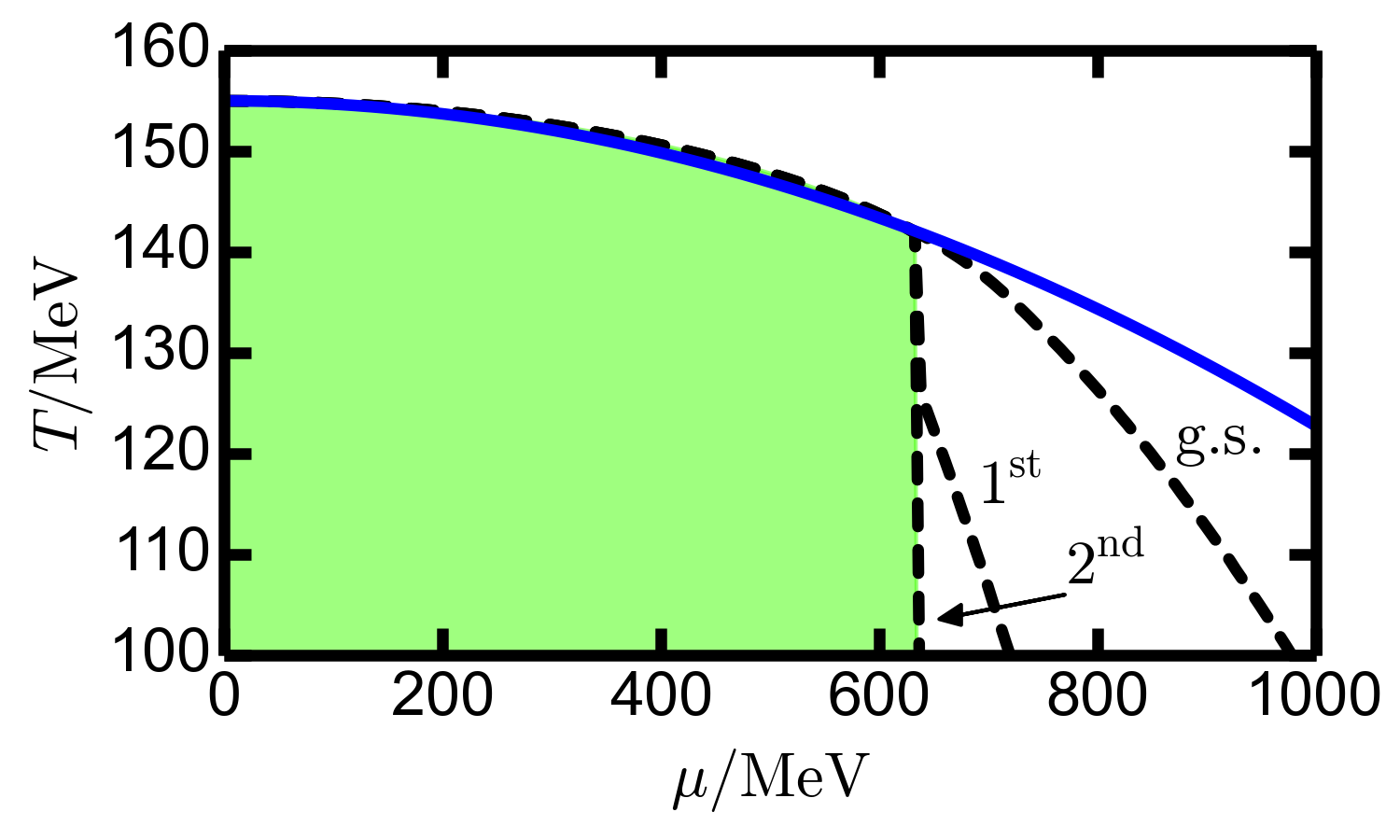} \includegraphics[scale=.49]{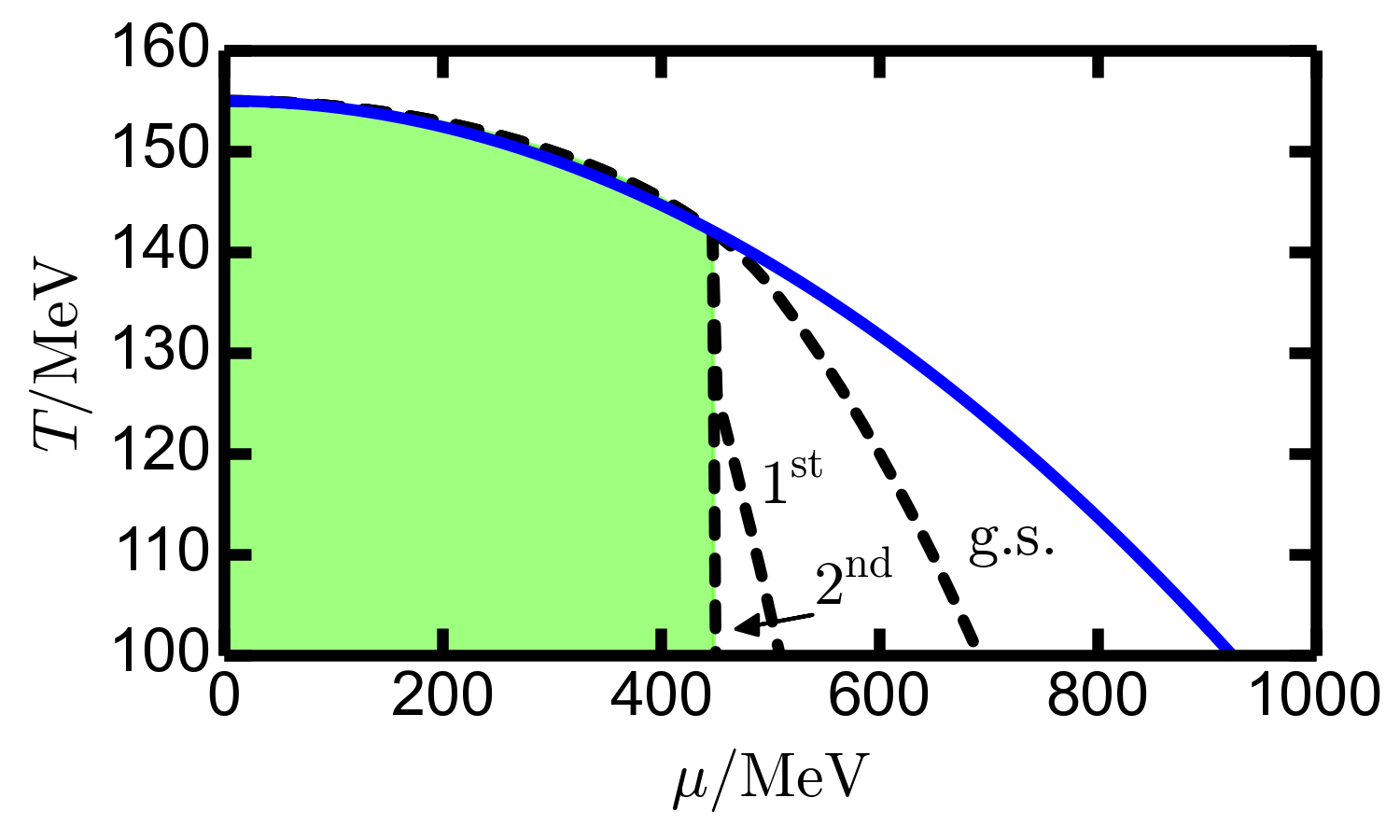}
 \caption{QCD phase diagram with two options of the freeze-out/cross-over curve (\ref{entwicklung}) (solid blue curves, left panel: $\kappa=0.005$, right panel: $\kappa=0.01$); note that (\ref{entwicklung}) without higher-order terms holds true only in the small-$\mu$ region.
 In the green areas, the thermal gas solution applies. Its upper boundary is given by $T_{\min}(\mu)$.
 The disappearance temperatures $T_{\dis}$ as a function of $\mu$ (dashed curves) of the first three vector meson states according to Eq. (\ref{schr}) with the potential (\ref{hotpot}) (parameters: $p=1.99$, $\tilde \mu =0.5$, $c=443$ MeV (cf. set 2.0 of \cite{ich}), $\tilde T_{\min}=155$ MeV, $c\tilde z_{\min}=2$) are adjusted by $\gamma=7.85$ (left) and $\gamma=5.55$ (right).
 Up to $\mu=620$ MeV (left) or $\mu=440$ MeV (right) all states disappear instantaneously at $T=T_{\min}$. For larger values of $\mu$, where only the black-hole solution is valid (white regions), the third and all higher states do not exist at all (indicated by the vertical dashed lines); the ground state and the first excited state disappear sequentially.} \label{abb1} } 
\end{figure}

\begin{figure}
 \cen{\includegraphics[width=6.2cm, height=3.8cm]{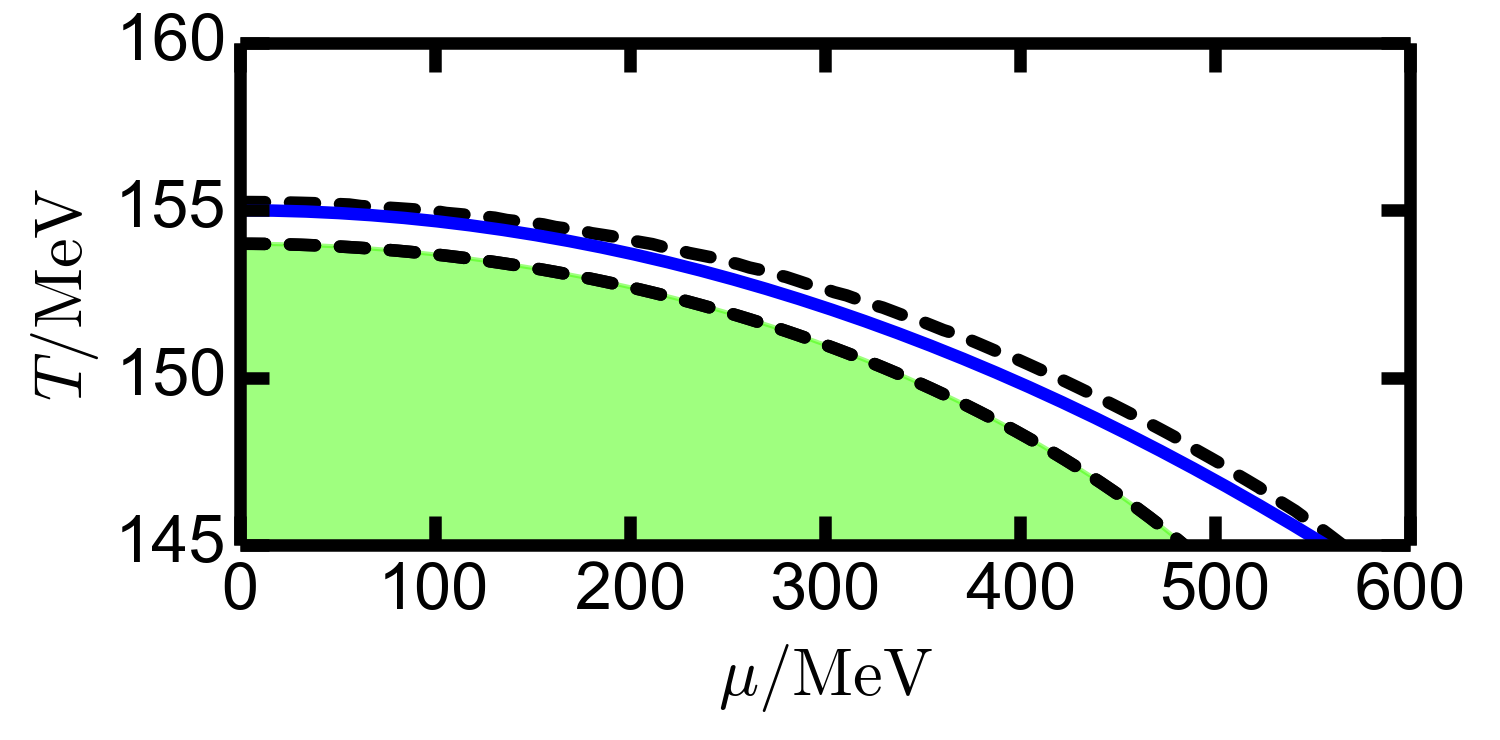} \includegraphics[width=6.2cm, height=3.8cm]{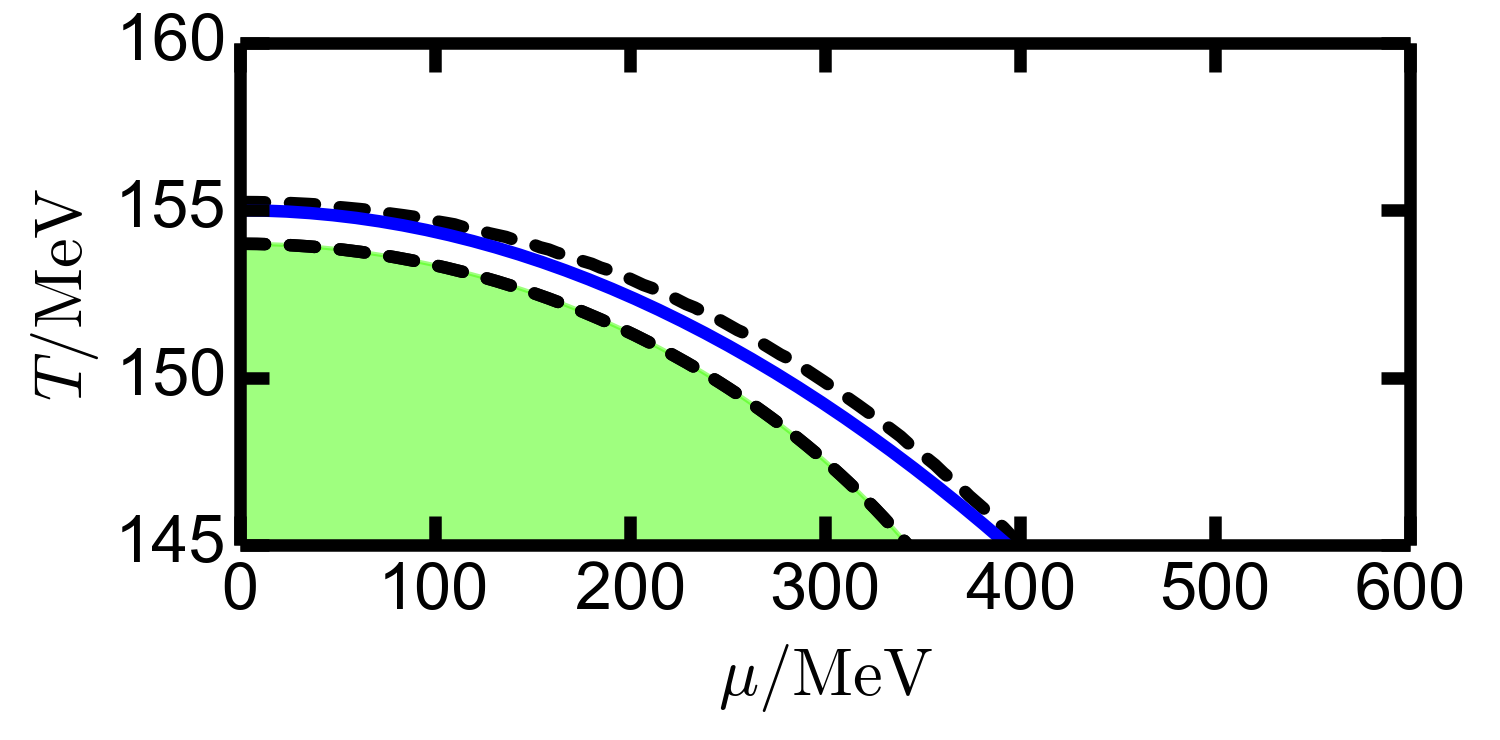}
 \caption{As Fig.~\ref{abb1} but for $\tilde T_{\min}=154$ MeV, $c\tilde z_{\min}=2.5$, $\gamma=8.79$ (left) and $\gamma = 6.22$ (right). For all values of $\mu$ the ground state disappears at a higher temperature than the radial excitations. While the thermal gas solution is valid, all excited states disappear instantaneously.} \label{abb2}}
\end{figure}

Depending on $\mu$, $\tilde T_{\min}$ and $\tilde z_{\min}$, $T(z_H)$ can display a minimum of $T_{\min}$ at $z_{\min}$ which translates into $T_{\min}(\mu)$. If so, then (\ref{f}) must be replaced by the trivial, non-black hole function $f=1$ for all $T<T_{\min}$, i.e. due to the Hawking-Page transition the thermal gas solution is the stable configuration. 
What remains is a selection of parameters $\tilde T_{\min}$, $\tilde z_{\min}$ and $\gamma$ to achieve $T_{\dis}^{\rm g.s.} (\mu) \cong T_{\rm f.o.}(\mu) \cong T_c(\mu)$. We take the leading order shape
  \begin{equation}
    T_{\rm f.o.}(\mu) \cong T_c(\mu) \cong T_0 \li(1- \kappa \li(\frac{\mu}{T_0} \ri)^2 + \ldots \ri) \label{entwicklung}
  \end{equation}
with $\kappa = 0.005 \ldots 0.01$ from \cite{Bazavov} (cf. also \cite{1605.09694}) and put for simplicity $T_0=T_c(\mu=0) = 155$ MeV without an error band. \\
The dependence of $T_{\dis}^{\rm g.s.}$ follows from numerical solutions of (\ref{schr}) with the potential (\ref{hotpot}), where the $\mu$ dependence comes from (\ref{f}, \ref{T}). We employ here the parameters $p=1.99$, $\tilde \mu=0.5$ and $c=443$ MeV which provide one possible setting of a Regge trajectory with $\alpha=0.71$ GeV$^2$ and $\beta=0.75$ GeV$^2$ at $T=\mu=0$, as shown in \cite{ich}. The particular choice $\tilde T_{\min}=155$ MeV and $c\tilde z_{\min}=2$ is for a scenario where, for $\mu=0$, the thermal gas solution is valid for all temperatures $T<\tilde T_{\min}$. That is, for $T<\tilde T_{\min}$ the vector meson spectrum is as at $T=0$ with the implication that the thermo-statistical model analysis applies in that region with standard vacuum masses of hadrons. At $T>\tilde T_{\min}$, however, the black hole solution must be accomplished. Equation (\ref{schr}) does not allow for normalisable solutions at $T>\tilde T_{\min}$, i.e.~just at $T=\tilde T_{\min}$ the hadron states (here shown only for vector mesons) disappear. In such a special setting one therefore identifies both the (chiral) cross over point and the chemical freeze-out temperature at $\mu=0$ with (de)confinement. We adjust the remaining parameter $\gamma$ such to put the disappearance temperature of the ground state, $T_{\dis}^{\rm g.s.}(\mu)$ (upper dashed curve), on the freeze-out/cross-over curve (solid blue curve) in the parametrisation (\ref{entwicklung}). Using the above quoted values of curvature measure $\kappa$ in the spirit of upper and lower bounds we find the results exhibited in Fig.~\ref{abb1}. Up to a certain critical value of the chemical potential the disappearance curve of the lowest vector meson states is on the top of the freeze-out/cross-over curve for a given value of $\kappa$. The related physical interpretation is that once a cooling piece of deconfinement matter reaches $T_{\dis}^{\rm g.s., 1^{\rm st}, \ldots}(\mu)$, it hadronizes by occupying statistically the available hadron states. \\
The above sketched scenario can be relaxed, by minor parameter variations, to have $T_{\dis}^{\rm g.s.}>T_{\dis}^{\rm 1^{st}} > T_{\dis}^{\rm 2^{nd}} \ldots$, i.e.~ a sequential appearance of vector meson states upon cooling. Figure \ref{abb2} exhibits a possibility where the first two states appear sequentially in a narrow corridor centred at $T_c(\mu)$ for small $\mu$. If such a behaviour can be established for other hadron species too, it is still consistent with the application of the thermo-statistical models. \\
It is premature to extrapolate the described scenario to too large values of $\mu$, and thus to critical point issues, since (i) Eq.~(\ref{entwicklung}) relies on the leading-order term and (ii) lacking knowledge on $T_c(\mu)$, i.e.~whether $T_{\rm f.o.}(\mu) \cong T_c(\mu)$ at larger values of $\mu$, and (iii) unsettled options in constructing other blackness functions beyond (\ref{f},\ref{T}). 

\section{Summary} \label{sec4}
The famous soft-wall model \cite{KKSS} represents a particular realisation of ideas anchored in the AdS/CFT correspondence. It can be modified to accommodate a Regge type spectrum a radial excitations of vector mesons. Considering vector mesons as prototypical representatives of hadrons one can further modify such a gravity field duality model to study the fate of certain hadron species immersed in a hot and dense ambient medium. Parameters can be tuned to let disappear vector mesons as normalisable modes above a temperature to be identified tentatively with ``deconfinement temperature'' or, more specifically, with the chiral cross-over temperature $T_c$ \cite{ich}, thus extending the approach in \cite{Colangelo}. Following, e.g. \cite{Bazavov} (see also \cite{pbm}) in identifying the chemical potential dependence of $T_c(\mu)$ with the freeze-out systematics found from heavy-ion experiments and condensed in $T_{\rm f.o.}(\mu)$ at small $\mu$, we have demonstrated that the suitably adopted soft-wall model allows for a consistent scenario: Once a temperature $T_{\dis}$ is reached upon cooling of a piece of ``deconfined matter'', hadrons appear, either suddenly at once or sequentially in a narrow corridor of temperatures, and are ready for statistical distribution. 

\begin{appendix}
\section{} \label{anhang}
The goal is to extend the black-hole function in AdS, $f_{\rm BH} (z) = 1-(z/z_H)^4$, yielding $T_{\rm BH} (z_H) = 1/(\pi z_H)$, and the Reissner-Nordstr\"om black-hole function, $f_{\rm RN}(z) =  1-(1+\frac12 \hat \mu^2)(z/z_H)^4 + \frac12 \hat \mu^2 (z/z_H)^6$ in AdS, yielding $T_{\rm RN} (z_H) = (\pi z_H)^{-1} (1-\hat \mu^2)$ \cite{Erdmenger}. Clearly, $f_{\rm RN}(z;z_H,\hat \mu=0)=f_{\rm BH}(z;z_H)$. As in \cite{ich} we start from the general statement that for all positive $i$ with $i>4(\pi z_HT(z_H)-1)=:4r$ the function $f$ defined by
  \begin{equation}
   f(z) = 1- \frac{z^4}{z_H^4} \li(1+ \frac{4r}{i}  \li(\frac{z^i}{z_H^i}-1\ri) \ri)
  \end{equation}
is a suitable blackness function, i.e. $f(z=0,z_H) = 1$, $(\Dp^i_z f)_{z \to 0} =0 $ for $i=1,2,3$ and the simple zero at the horizon, $f(z=z_H;z_H)=0$. To recover the Reissner-Nordstr\"om case, we observe that $r=-\hat\mu^2$ and $i=2$ are required. To construct a proper blackness function we can apply any function $h: \R \to \R$ with $h$ positive, $h(x)>x$ for all $x \in\R$ and $h(-4\hat \mu^2)=2$ and set $i=h(4r)$. One possibility is $h(x) = 2 e^{ax+4\hat\mu^2}$ for all $a \geq 1/2e$ which yields (\ref{f}) for $a=1/2e$.

\end{appendix}

\end{document}